\documentclass [aip,
reprint,%
]{revtex4-1}

\usepackage{graphicx}%
\usepackage{dcolumn}%
\usepackage{bm} 

\usepackage{multirow}
\usepackage{hyperref} 
\usepackage{makecell} 
\usepackage{amssymb} 
\usepackage{amsmath,mathtools}
\usepackage{wasysym} 
\usepackage[makeroom]{cancel} 
\usepackage{soul} 
\usepackage{xcolor}
\usepackage[flushleft]{threeparttable} 
\usepackage{booktabs}  
\usepackage[T1]{fontenc} 
\usepackage{arydshln} 
\usepackage{gensymb} 

\usepackage{nomencl} 
\makenomenclature
\usepackage{etoolbox}
\renewcommand\nomgroup[1]{%
	\item[\bfseries
	\ifstrequal{#1}{A}{Roman}{%
		\ifstrequal{#1}{B}{Greek}{%
			\ifstrequal{#1}{C}{Subscripts}{}}}%
	]}

\usepackage{lineno} 
\usepackage{stackengine}
\stackMath

\newcommand{\etal}{\textit{et al}.}
\newcommand{\ra}[1]{\renewcommand{\arraystretch}{#1}}

\begin {document}

\title{Evolution of electron cross-field transport induced by an equilibrium azimuthal electric field in an $\mathbf{E}\times\mathbf{B}$ Hall thruster discharge under an azimuthally inhomogeneous neutral supply} 

\author{J. Bak}
\email{j.bak@al.t.u-tokyo.ac.jp; junhwib@tamu.edu}
\altaffiliation[Currently at ]{Aerospace Engineering Department, Texas A\&M University.}

\author{R. Kawashima}%
\affiliation{Graduate School of Engineering, The University of Tokyo, Tokyo 113-8656, Japan}

\author{J. Simmonds}
\affiliation{Mechanical and Aerospace Engineering, Princeton University, Princeton, New Jersey 08543, USA}

\author{K. Komurasaki}
\affiliation{Graduate School of Engineering, The University of Tokyo, Tokyo 113-8656, Japan}

\date{10 September 2021}

\begin{abstract}
	The electron cross-field transport by the induced azimuthal electric field in a Hall thruster exhibits the mobility scaled by $1/B$. This study investigates parameters affecting this transport over a Hall thruster's distinct regions, such as the ionization, acceleration, and plume region. The main focus is on the nonzero equilibrium azimuthal electric field induced by an azimuthally inhomogeneous neutral supply. A fast Fourier transform analysis of the plasma structure reveals that the wavenumber $k$ of the azimuthal plasma structure increases from $k=2$, which is the input condition, to $k=4$ in the plume region, and that the total axial flux caused by the azimuthal electric field is mainly induced from the structures of the dominant Fourier components. The azimuthal phase relation between plasma potential and density is formed to maximize the axial electron flux at the plume region and starts varying along with other plasma properties as electrons flow toward the acceleration region. The spatial evolution of the effective axial mobility coefficient is extracted, and its regional characteristics are discussed.
	
\end{abstract}

\maketitle 
\section{Introduction}
Electron transport across magnetic fields in various types of $\mathrm{E} \times \mathrm{B}$ plasma equipment\cite{Rossnagel1987,Martines2001,Quraishi2002,Morozov1972,Hagelaar2003} often shows anomalously higher mobility than classical electron-neutral collisional mobility. The measured mobility, which scales as $1/B$, is commonly modeled with a spatially varying coefficient\cite{Rossnagel1987, Boeuf1998, Keidar2004, Hagelaar2003} to reproduce experimentally obtained results. This anomalous mobility has been theoretically shown since the early 1960s to result from plasma density fluctuations.\cite{Yoshikawa1962} Later, Janes and Lowder\cite{Janes1966} measured fluctuating density and potential in a plasma accelerator, with an external radial $B$ field and an applied axial  $E$ field, and showed that an azimuthal correlation of $n_\mathrm{e}$ and the induced $E_\theta$ exists and results in a net anomalous cross-field diffusion. 

In the common cylindrical $r$-$\theta$-$z$ coordinate system of Hall thrusters (HTs) with the radial magnetic field and axial electric field, the net axial cross-field electron flux $\Gamma_{\mathrm{e}z, E_\theta}$ by the induced azimuthal electric field  can be expressed by 
\begin{eqnarray}
	\Gamma_{\mathrm{e}z, E_\theta}^- &=& \frac{1}{B_r} \left\langle  n_\mathrm{e} E_\theta\right\rangle, \label{eq:gam_ez}
\end{eqnarray}
where the azimuthal average $\left\langle X \right\rangle = \frac{1}{2\pi} \int_{0}^{2 \pi} X d\theta $ with an arbitrary variable $X$, and the superscript of the minus sign denotes the $-z$ direction. Such cross-field transport from the correlation of $n_\mathrm{e}$ and the induced $E_\theta$ in  $\mathbf{E} \times \mathbf{B}$ plasmas is reported in a number of numerical\cite{Hirakawa1995,Lafleur2016,Croes2017,Carlsson2018} and experimental studies\cite{Meezan2001,Ellison2012,Bak2019,Bak2020} of $\mathrm{E} \times \mathrm{B}$ plasma. Similar transport effects by an induced (fluctuating) field is often discussed in the magnetically confined fusion machines\cite{Birkenmeier2011}.

In Hall thrusters, research on cross-field electron transport shows that the physics behind the axial enhancement of transport is commonly due to azimuthal mechanisms. For example, azimuthally propagating plasma instabilities have been widely studied, and the corresponding enhancement of electron transport by the fluctuating $E_\theta$ are reported in multiple frequency ranges, i.e., gradient drift instability or Simon-Hoh instability in the kHz order\cite{Janes1966, Ellison2012, Boeuf2019}, and the electron drift instability in the MHz order\cite{Lafleur2017, Croes2017, Boeuf2018}. Enhanced electron transport has also been reported during azimuthal nonuniform propellant supply.\cite{Fukushima2009a,Ding2019a,Bak2019} In our past work\cite{Bak2019}, we experimentally observed a non-zero equilibrium $E_\theta$. Electron transport by the induced $E_\theta$ in high Hall parameter $\mathrm{E} \times \mathrm{B}$ plasmas provides mobility that is 2-3 orders of magnitude higher than the classical mobility, and becomes the dominant cross-field transport mechanism which causes the effective mobility to be proportional to 1/B.\cite{Bak2019}

Even though the direct evaluation of Eq.~(\ref{eq:gam_ez}) has been made and has shown the transport effect of the induced field for various $E_\theta$ sources\cite{Ellison2012, Lafleur2016, Croes2017, Boeuf2018, Bak2019}, detailed studies focusing on affecting parameters such as an azimuthal phase difference between plasma density and potential, which can be critical to yield the net cross-field transport, have been lacking. For experimental approaches, most studies focused on regimes where the phase relation between the plasma density and potential in the spokes were completely in-phase \cite{Janes1966}, or not completely in-phase\cite{Ellison2012, Skoutnev2018}. However, these observations were limited to one specific axial location. Thus, a study to understand how transport by $E_\theta$ evolves throughout the wide characteristic regions between anode and cathode has been lacking.
\begin{figure}
	\centering
	\includegraphics[width=0.99\linewidth]{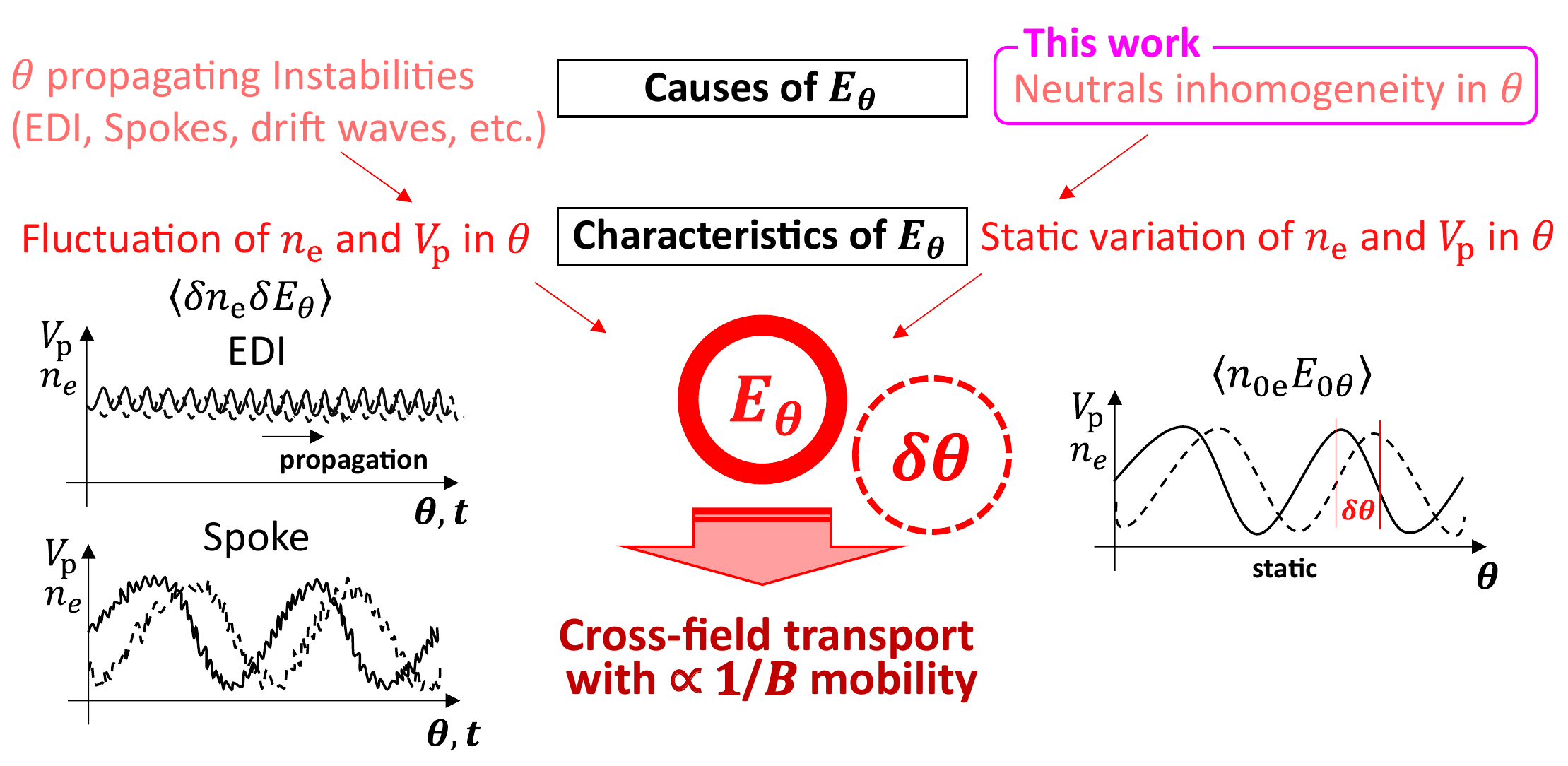}
	\caption{Diagram showing various causes of $E_\theta$ and typical spatio-temporal characteristics of plasma density $n_\mathrm{e}$ and potential $V_\mathrm{p}$. When there exists an azimuthal phase difference $\delta \theta$ between $n_\mathrm{e}$ and $V_\mathrm{p}$, $E_\theta$ leads to the cross-field transport.}
	\label{fig:ethetrans}
\end{figure}

In this work, we aim to further reveal details of the transport induced by $E_\theta$. It should be noted that because the used experimental data of plasma properties are time-averaged values and each property is independently diagnosed, our consideration is solely focused on the transport resulting from the zeroth-order equilibrium distributions of plasma properties in the $z-\theta$ plane. The $E_\theta$ here is induced by the azimuthally inhomogeneous neutral particles, and is a fundamentally different source of $E_\theta$ compared to ones induced by the azimuthally propagating instabilities\cite{Janes1966, Ellison2012, Carlsson2018, Lafleur2016, Croes2017}. Thus, when the \textit{total} axial electron flux ${\Gamma}_{\mathrm{e}z\mathrm{,tot}}^-$ is expressed as a sum of the flux from the static equilibrium variation $\bar{\Gamma}_{\mathrm{e}z}^-$ and that from fluctuations $\tilde{\Gamma}_{\mathrm{e}z}^-$, thus ${\Gamma}_{\mathrm{e}z\mathrm{,tot}}^- =  \bar{\Gamma}_{\mathrm{e}z}^- + \tilde{\Gamma}_{\mathrm{e}z}^-$, our discussion is focused on the time-independent flux component $\bar{\Gamma}_{\mathrm{e}z}^-$. We omit the over-bar for the simple notation in the rest of the paper. Any transport effects resulting from such fluctuating properties are not taken into account; we leave the study of transport effects from time-varying, multi-scale mechanisms for future work.

The various causes of $E_\theta$, the typical spatiotemporal characteristics of $n_\mathrm{e}$ and $V_\mathrm{p}$, and the position of the current work are diagrammed in Fig.~\ref{fig:ethetrans}. Note that the diagram is simplified with a specific main source. Strictly speaking, any structure in a steady-state discharge would not be purely from one source but from every interaction of physics happening in the discharge, including second-order mechanisms such as coupling of instabilities and gas/electron/ion flows. Even though it is unknown how such second-order physics influence the final structure, the equilibrium structure in the current work provides experimentally accessible azimuthal variations of plasma properties to investigate the transport by $E_\theta$.
It is noteworthy that the $1/B$ form of cross-field transport is not limited to the Bohm type. Even though the fundamental cause of $E_\theta$ is different, the present work is expected to provide fruitful information on understanding the transport induced by $E_\theta$.  Also, the unique structural features from the nonuniform operation can be useful to validate numerical models developed for azimuthal physics.\cite{Escobar2015, Croes2017, Chernyshev2019, Kawashima2017, Kawashima2018, Kawashima2021}

Transport effects by sub-structures are studied, and the dominant structure mode is identified. Also, the spatial evolution of individual parameters affecting $\Gamma_{\mathrm{e}z, E_\theta}^-$ is investigated over a wide axial region which includes all characteristics regions of Hall thrusters (near-anode, acceleration, and plume region). The changes in these parameters are explained using the force balance on electrons. It is noteworthy that the present work is distinguished from our past works where the focus was made on resolving the plasma structure and identifying electron flux by $E_\theta$,\cite{Bak2019} and characterizing discharge behavior depending on the input neutral inhomogeneity level\cite{Bak2020}. 

The structure of this paper is as follows. In Sec.~\ref{sec:paraset}, we define a set of parameters affecting the transport induced by $E_\theta$. In Sec.~\ref{sec:result_pla_str}, plasma structure obtained during the operation with the azimuthally modulated input neutrals is provided and discussed; the experimental setup and plasma data (Sec.~\ref{sec:data}) and the wavenumber analysis of plasma structure (Sec.~\ref{sec:wavnum}). In Sec.~\ref{sec:result_Gam_Ethe}, regional characteristics of $\Gamma_{\mathrm{e}z, E_\theta}^-$ are analyzed and discussed; evolutions of individual affecting parameters (Sec.~\ref{sec:evol}), the relationship between the parameters (Sec.~\ref{sec:forbal}), and the effective mobility coefficient (Sec.~\ref{sec:mobil}). Lastly, we summarize the work in Sec.~\ref{sec:sum}.

\section{A set of parameters affecting $\Gamma_{\mathrm{e}z, E_\theta}^-$}\label{sec:paraset}
Assuming azimuthal distributions of electron density  $n_\mathrm{e}$ and plasma potential $V_\mathrm{p}$ of wavenumber $k$ in sinusoidal forms: $n_\mathrm{e} = n_\mathrm{e0} + n_\mathrm{e1} \sin(k\theta)$ and $V_\mathrm{p} = V_\mathrm{p0} + V_\mathrm{p1} \sin(k(\theta+\delta \theta))$, where $\delta \theta$ is the azimuthal phase difference between $n_\mathrm{e}$ and $V_\mathrm{p}$, Eq.~(\ref{eq:gam_ez}) can be re-expressed as
\begin{eqnarray}
	\Gamma_{\mathrm{e}z, E_\theta}^- &=& 0.5\sin{(k\delta\theta)}\cdot {n_\mathrm{e1}}\cdot{E_{\theta\mathrm{1}}}\cdot\frac{1}{B_r}, \label{eq:gam_ez_}
\end{eqnarray}
which shows the set of parameters affecting the flux generated by $E_\theta$: 
\begin{itemize}
	\item $0.5 \sin{(k \delta \theta)}$ : the effective weight coefficient from the phase difference between $n_\mathrm{e}$ and $V_\mathrm{p}$. The phase difference which maximizes $\Gamma_{\mathrm{e}z, E_\theta}^-$ is then $\delta \theta^* \equiv \frac{\pi}{2k}$.
	\item  $n_\mathrm{e1}$ : the amplitude of the azimuthal electron density waveform.
	\item  $E_{\theta\mathrm{1}}$ : the amplitude of the azimuthal electric field waveform given as $E_{\theta\mathrm{1}} = k V_\mathrm{p1}/{R_\mathrm{ch}} $, where $R_\mathrm{ch}$ is the mean discharge channel radius.
	\item  $B_r$ : the magnetic flux density.
\end{itemize}

\section{Plasma structure on the $z$-$\theta$ plane}\label{sec:result_pla_str}
\subsection{Experimental setup and available properties}\label{sec:data}

\begin{figure}
	\centering
	\includegraphics[width=0.99\linewidth]{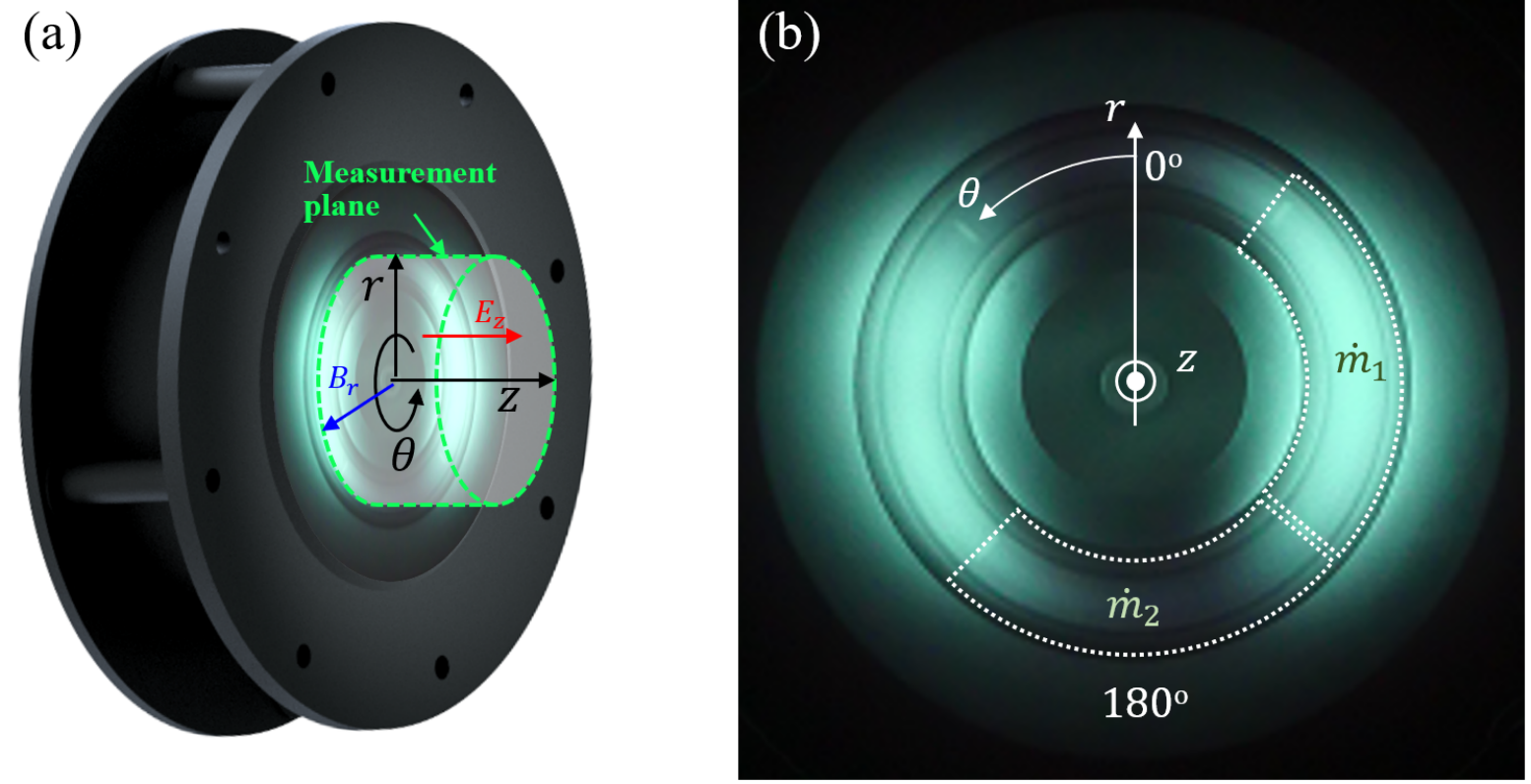}
	\caption{(a) Coordination system and $z$-$\theta$ measurement plane along the channel center. (b) Thruster operation image with azimuthal nonuniform propellant injection (mass flow rate $\dot{m}_1 > \dot{m}_2$).}
	\label{fig:setup}
\end{figure}

\begin{figure*}
	\centering
	\includegraphics[width=12 cm]{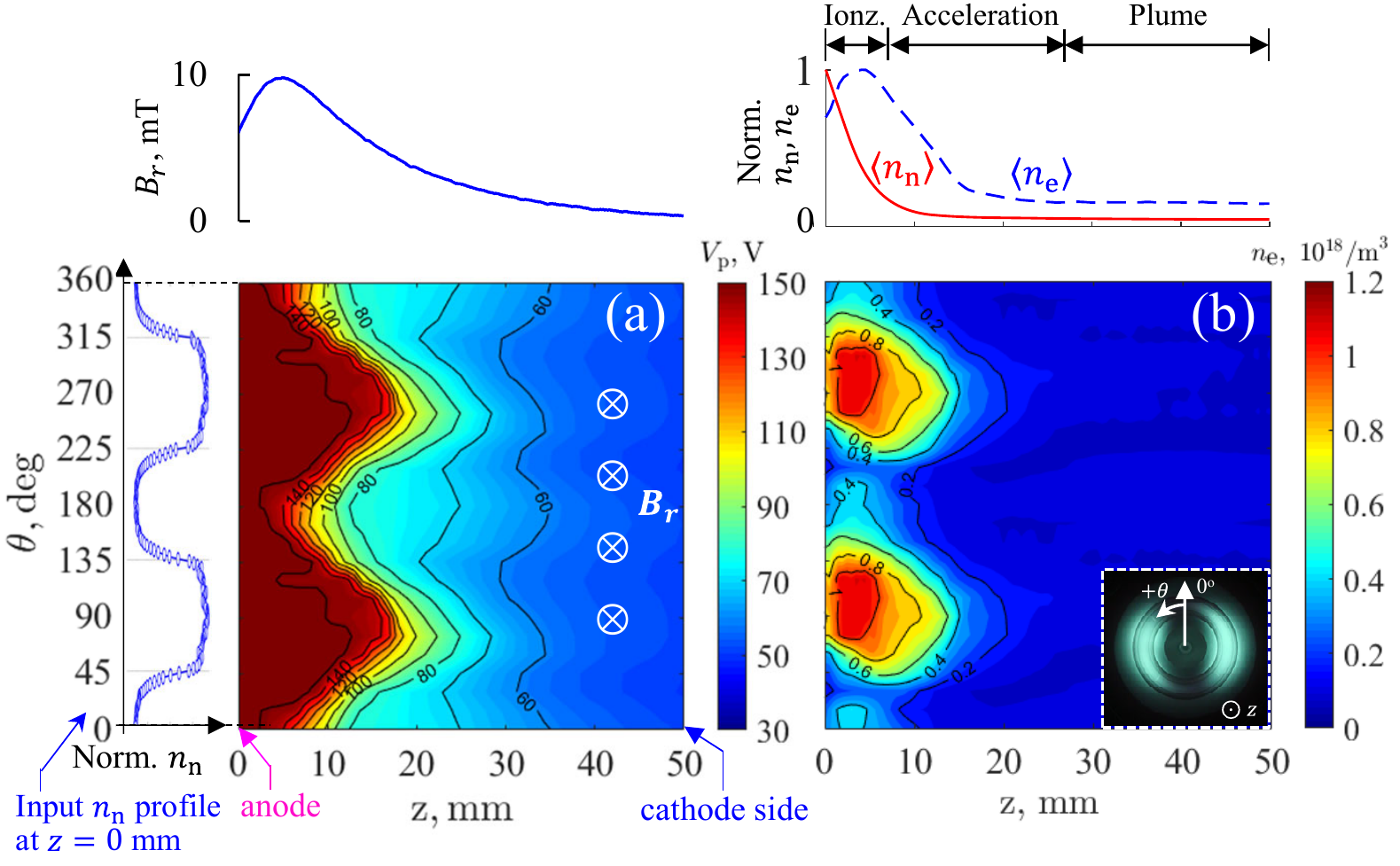}
	\caption{(a) The plasma potential $V_\mathrm{p}$ and (b) the electron density $n_\mathrm{e}$ map on the $z-\theta$ plane of a Hall thruster operated with azimuthally non-uniform neutral particles. In $V_\mathrm{p}$ map, the input neutral density $n_\mathrm{n}(\theta)$ (normalized by the max value) is on the left, and the radial magnetic flux density $B_r(z)$ is shown above. Azimuthally averaged $n_\mathrm{n}$ and $n_\mathrm{e}$ in the angle brackets, normalized by the its maximum value are shown above the $n_\mathrm{e}$ map. The discharge image of the thruster is shown in a box at the bottom right. Three characteristics regions are indicated; ionization ($z$ = 0 - 7 mm), acceleration ($z$ = 7 - 25 mm), and plume ($z$ > 25 mm). Reprinted with permission from Bak et al., Physics of Plasmas \textbf{26}, 073505 (2019)}
	\label{fig:vpnedis}
\end{figure*}

Our analysis in the current work is based on the experimental data of plasma properties obtained in Ref.~\onlinecite{Bak2019}. The data were obtained on a cylindrical Hall thruster $\mathrm{E} \times \mathrm{B}$ discharge, Fig.~\ref{fig:setup}(a), in which the plasma features a stationary azimuthal modulation due to nonuniform propellant injection, Fig.~\ref{fig:setup}(b). In Ref.~\onlinecite{Bak2019}, propellant was fed by two mass flow controllers having flow rate $\dot{m}_1$ and $\dot{m}_2$ respectively. The nonuniformity level of the supplied propellant is defined as ${\Xi} = \dot{m}_\mathrm{diff}/\dot{m}_\mathrm{tot}$, where $\dot{m}_\mathrm{diff}$ is the difference of the mass flow rates in the two flow controllers, and $\dot{m}_\mathrm{tot}$ is the total mass flow rate supplied. The thruster was operated with $\Xi$=0.8 and $\dot{m}_\mathrm{tot}$=$2.04$ mg/s at the discharge voltage $150$ V while the background pressure was maintained below $6.6$ mPa. 

The measured $z$-$\theta$ surface was along the center of the channel at $r$=29 mm. Probes were inserted axially into the thruster at azimuthal positions between 60\degree~and 240\degree~ with 15\degree~ increments, which was between an axis of symmetry. The remaining half of the thruster was assumed to be identical, and data showing the remaining portion has been duplicated from the former portion for visual clarity. The sampling rate for each axial insertion was 100 kS/s, giving > 200 samples per mm. A total of two to four axial profiles at each azimuthal location were measured and averaged. The hot emissive and cold floating probes were used for diagnostics of plasma potential $V_\mathrm{p}$ and electron temperature $T_\mathrm{e}$ (the standard random error < 3\% and systematic error < 10\%), and a biased single Langmuir probe was used to obtain the plasma density $n_\mathrm{e}$ (the standard random error < 5\% at $z<25$ mm and < 20\% at $z>25$ mm, and the systematic error < 23\%). Please note that at the operation condition, the discharge current oscillation, defined as the ratio of the root mean square of the discharge current to the mean discharge current, was < 2\%, and the corresponding fluctuation of plasma properties was also at a similar level.

The neutral density $n_\mathrm{n}$ was obtained by numerically solving the continuity equation with the experimentally obtained plasma parameters and the input profile calculated at $z=0$ mm by a 3D particle calculation on the anode assembly. Please refer to Ref.~\onlinecite{Bak2019} for further details of the calculation domain and method.

Figure~\ref{fig:vpnedis} shows the plasma potential map (a) and the electron density map (b) on the $z$-$\theta$ plane taken from Ref.~\onlinecite{Bak2019}. The plasma structure results from the azimuthally inhomogeneous input of neutrals. On the left side of Fig.~\ref{fig:vpnedis}(a), the input $n_\mathrm{n}$ profile in the azimuth is shown, and on the top, $B_r(z)$ is shown. On the top of the $n_\mathrm{e}$ map in Fig,~\ref{fig:vpnedis}(b), the normalized $n_\mathrm{n}$ and $n_\mathrm{e}$, which are averaged azimuthally, are shown as a function of axial location. Based on physical properties, we categorize three characteristics regions; ionization ($z$ = 0 - 7 mm), acceleration ($z$ = 7 - 25 mm), and plume ($z$ > 25 mm). Further details of the structures, diagnostics, and calculation can be found in Ref.~\onlinecite{Bak2019}.

In summary, four experimentally obtained properties (plasma properties $V_\mathrm{p}$, $T_\mathrm{e}$, $n_\mathrm{e}$ and the magnetic flux density $B_r$) and one numerically obtained $n_\mathrm{n}$ are primary available quantities in this work, and these are used on the post-calculation of other physical properties. 

\subsection{Wavenumber of plasma structure}\label{sec:wavnum}
\begin{figure}
	\centering
	\includegraphics[width=0.99\linewidth]{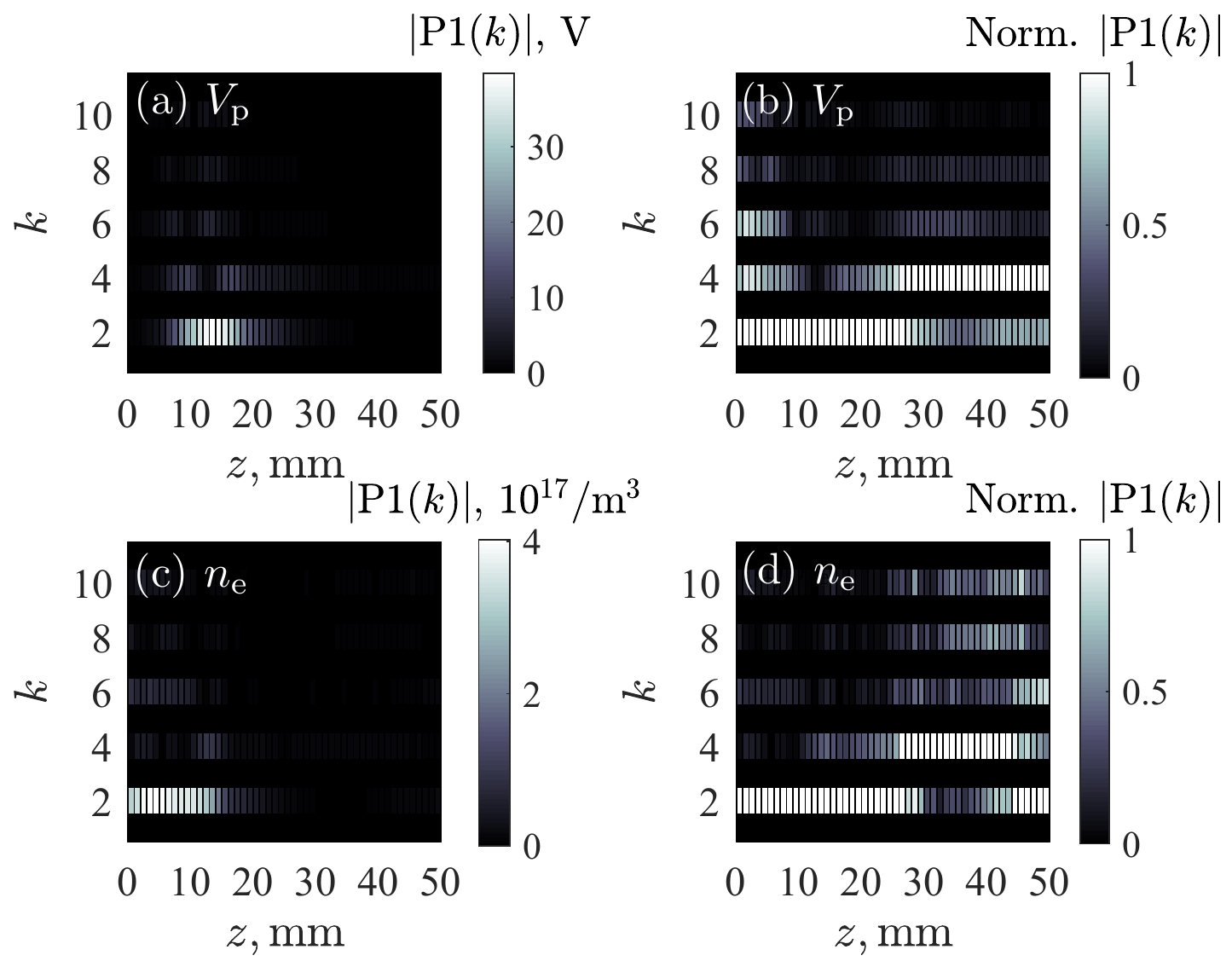}
	\caption{Maps of Fourier transformed amplitude $|\mathrm{P1}(k)|$ of azimuthal distribution for (a) $V_\mathrm{p}$, and (b) normalized $|\mathrm{P1}(k)|$ by the maximum value at each axial location. (c-d) for  $n_\mathrm{e}$.}
	\label{fig:fftknormap}
\end{figure}
\begin{figure*}
	\centering
	\includegraphics[width=14 cm]{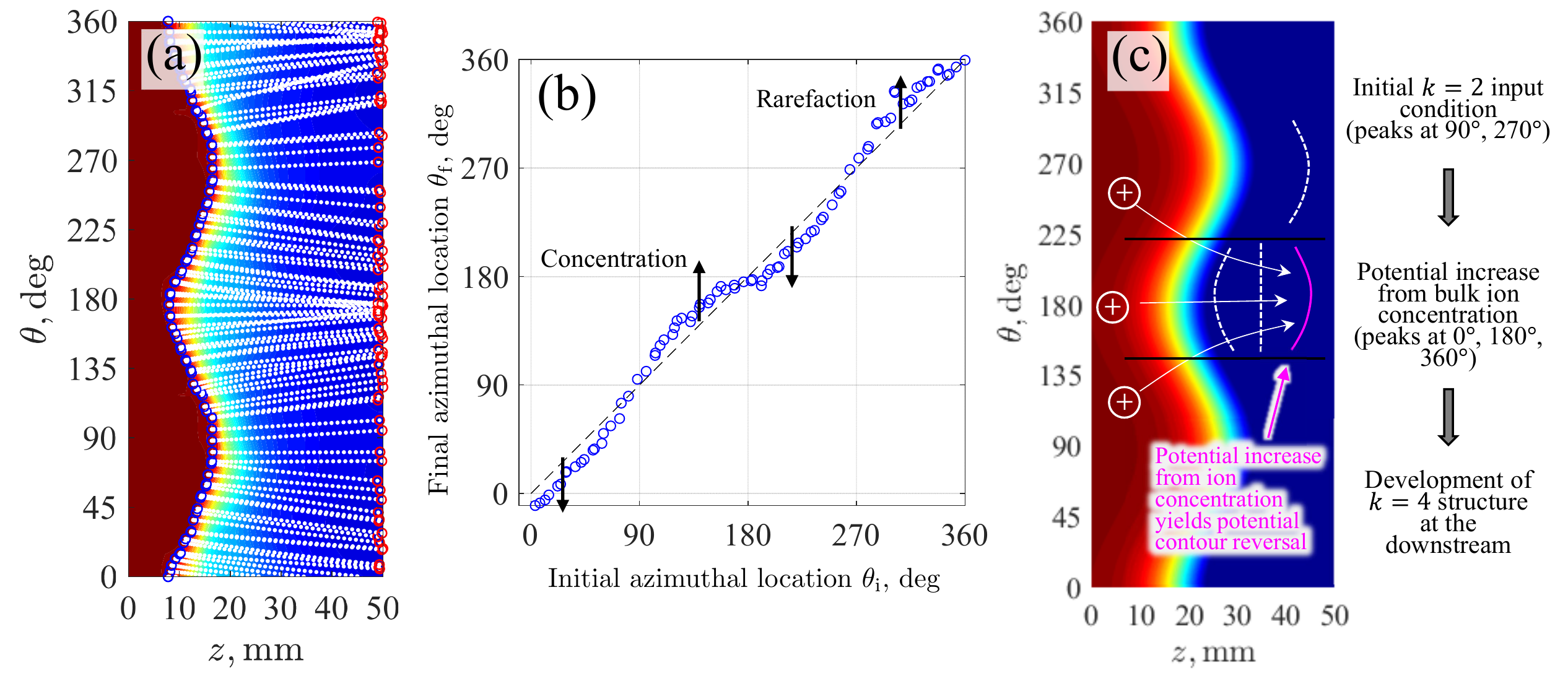}
	\caption{(a) Test particle shooting. Blue circles represent each test particle at initial locations which are along  $V_\mathrm{p}$=140 V contour, and red circles are their final location. White dotted lines represent trajectories. (b) Final arriving azimuthal location $\theta_\mathrm{f}$ with respect to initial azimuthal location $\theta_\mathrm{i}$. (c) Schematic of hypothesized structure formation at downstream.}
	\label{fig:iontrajec}
\end{figure*}
A fast Fourier transform (FFT), to extract the wavenumber $k$ of the plasma structure, is performed at each azimuthal distribution of the $V_\mathrm{p}$ and $n_\mathrm{e}$ at each axial location, and the results are shown in Fig.~\ref{fig:fftknormap}, which maps the single-sided amplitude $|\mathrm{P1}(k)|$ for $V_\mathrm{p}$ (Fig.~\ref{fig:fftknormap}(a)) and $n_\mathrm{e}$ (Fig.~\ref{fig:fftknormap}(c)). The intensity normalized by the maximum value at each axial location is shown in (Fig.~\ref{fig:fftknormap}(b)) and (Fig.~\ref{fig:fftknormap}(d)). The transition to the $k=4$ structure at $z\approx 25$ mm from the $k=2$ structure of the anode side stands out, implying that the influence of the neutral input condition, which imposes a constraint on forming the $k=2$ structure, becomes weak, and additional fine structures can form.

Because plasma potential can be closely related to the bulk ion (plasma density) distribution,\cite{Hagelaar2003} we investigate the ions' behavior under the formed structure. Considering the modulated input of neutral particles, we expect the ion production to be azimuthally modulated at the dense regions. This is indeed true when we see the plasma density structure. However, given the azimuthal plasma potential and density gradient, it can be assumed that these ions would quickly diffuse azimuthally from the main generation region to cover the channel as they follow the lower potential contours. As such, no modulation is considered for the particle shooting. Therefore, test ion particles are located along a contour of $V_\mathrm{p}$ =  140 V over $0\degree < \theta <360\degree $. An initial velocity of zero is given to the particles, and their trajectories are obtained by solving the motion equation (Fig.~\ref{fig:iontrajec}(a)). Blue markers indicate each test particle's origin. Their initial azimuthal location $\theta_\mathrm{i}$ and the final location $\theta_\mathrm{f}$ are plotted in Fig.~\ref{fig:iontrajec}(b). It is notable that ions' trajectories are mainly determined by the acceleration in the region of $k=2$ where the strong $E$ exists. Due to the potential structure in which the high potential area (at 90\degree and 270\degree) is shifted downstream because of the controlled neutral distributions, it is notable that ions concentrate at the potential valley (near $\theta=180\degree$).

Based on this result, we hypothesize that the concentration of these accelerated ions may lead to an additional high potential region, and in doing so, may contribute to the higher wavenumber structure at the downstream area, as described in Fig.~\ref{fig:iontrajec}(c). It is noteworthy that, however, this hypothesis may be valid only for a large inhomogeneity case. In the case of uniform gas injection, if we imagine decreasing the gas injection inhomogeneity little by little, the ion concentration/rarefaction would smooth out and reduce the high wavenumber structure. Additionally, it should be noted that the azimuthal location of the crests obtained here does not exactly match the experimental crest location. This may be because the used potential information is limited in the 2D $z$-$\theta$ plane, and so cannot take into account ions that could arrive at the same location from other radial positions in the experiment. A more accurate evaluation would require a full 3D experiment. Nonetheless, the result qualitatively shows that the bulk ions' trajectories can influence the finer downstream structure. 

\section{Analysis of $\Gamma_{\mathrm{e}z, E_\theta}^-$ on the $z$-$\theta$ plane}\label{sec:result_Gam_Ethe}

\subsection{Regional evolution of parameters affecting $\Gamma_{\mathrm{e}z, E_\theta}^-$}\label{sec:evol}

Even though the FFT results show the dominant Fourier components (high amplitude wavenumber mode), the amplitude itself cannot guarantee that the electron transport by such mode would also be dominant. This is because the effective weight coefficient is determined by the phase difference between $n_\mathrm{e}$ and $V_\mathrm{p}$ as seen in Eq.~(\ref{eq:gam_ez_}). 

\begin{figure}
	\centering
	\includegraphics[width=0.8\linewidth]{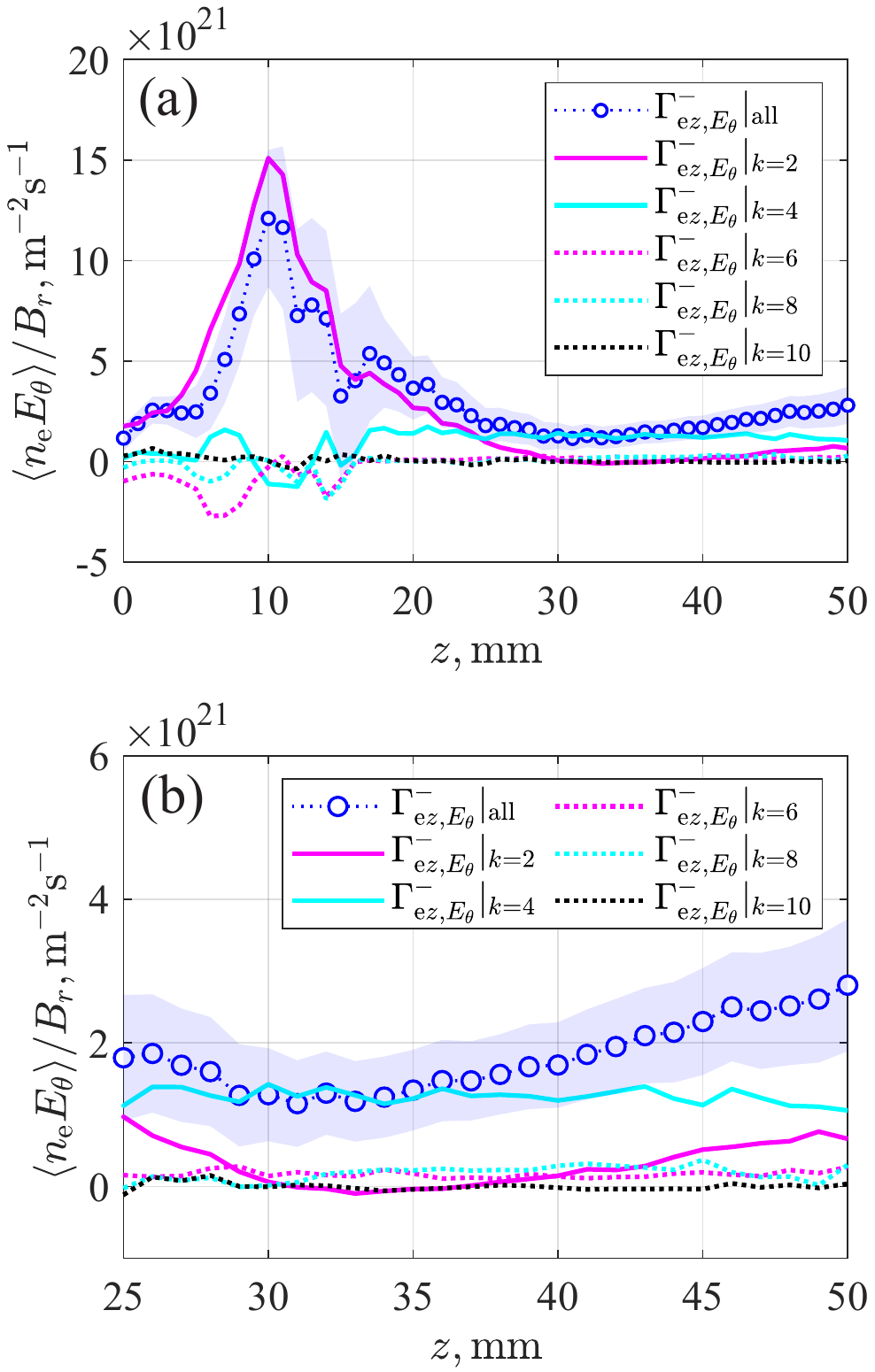}
	\caption{Comparison of the total axial flux $\Gamma_{\mathrm{e}z, E_\theta}^-$ (blue circle with the uncertainty shade band) and the fluxes from the Fourier components, $\Gamma_{\mathrm{e}z, E_\theta}^-$ for $k$ = 2, 4, 6, 8, and 10. The whole axial range ($z=$ 0 to 50 mm) is given in (a), and $z \ge 25$ mm are magnifed in (b). The dominant Fourier components, $\Gamma_{\mathrm{e}z, E_\theta}^-|_{k=2}$ for $z<25$ mm (solid magenta) and $\Gamma_{\mathrm{e}z, E_\theta}^-|_{k=4}$ for $z \ge 25$ mm (solid cyan) are notable.}
	\label{fig:fxcompa}
\end{figure}

To confirm whether the dominant Fourier components are responsible for the electron transport, the full axial flux by $E_\theta$ from the raw data, evaluated by Eq.~(\ref{eq:gam_ez}), is compared to the axial fluxes $\Gamma_{\mathrm{e}z, E_\theta}^-$ by the harmonics from $k=2$ to $k=10$, evaluated by Eq.~(\ref{eq:gam_ez_}). From Fig.~\ref{fig:fxcompa}, it can be seen that the total flux by $E_\theta$ $\Gamma_{\mathrm{e}z, E_\theta}^-|_\mathrm{all}$ are reasonably well represented by   $\Gamma_{\mathrm{e}z, E_\theta}^-|_{k=2}$ for $z<25$ mm (solid magenta) and $\Gamma_{\mathrm{e}z, E_\theta}^-|_{k=4}$ for $z \ge 25$ mm (solid cyan). Therefore, the flux by the structures of the dominant Fourier components are mainly responsible for the total flux. 

Based on the FFT analysis and the current result, for the simplicity of the analysis, the wavenumbers $k=2$ for $z<25$ mm and $k=4$ for $z \ge 25$ mm are chosen for the sine function fitting of the azimuthal distribution in form of $f(\theta) = a + b \sin(k(\theta+c))$ where $a$, $b$, and $c$ are the fitting coefficients. The fitting provides all affecting parameters to evaluate the axial electron flux $\Gamma_{\mathrm{e}z}^-$ given in Eq.~(\ref{eq:gam_ez_}). Fitting is done from $z=0$ mm to $z=50$ mm with 1 mm increments. More details for the fitting can be found in Appendix~\ref{sec:app}.

\begin{figure*}
	\centering
	\includegraphics[width=12 cm]{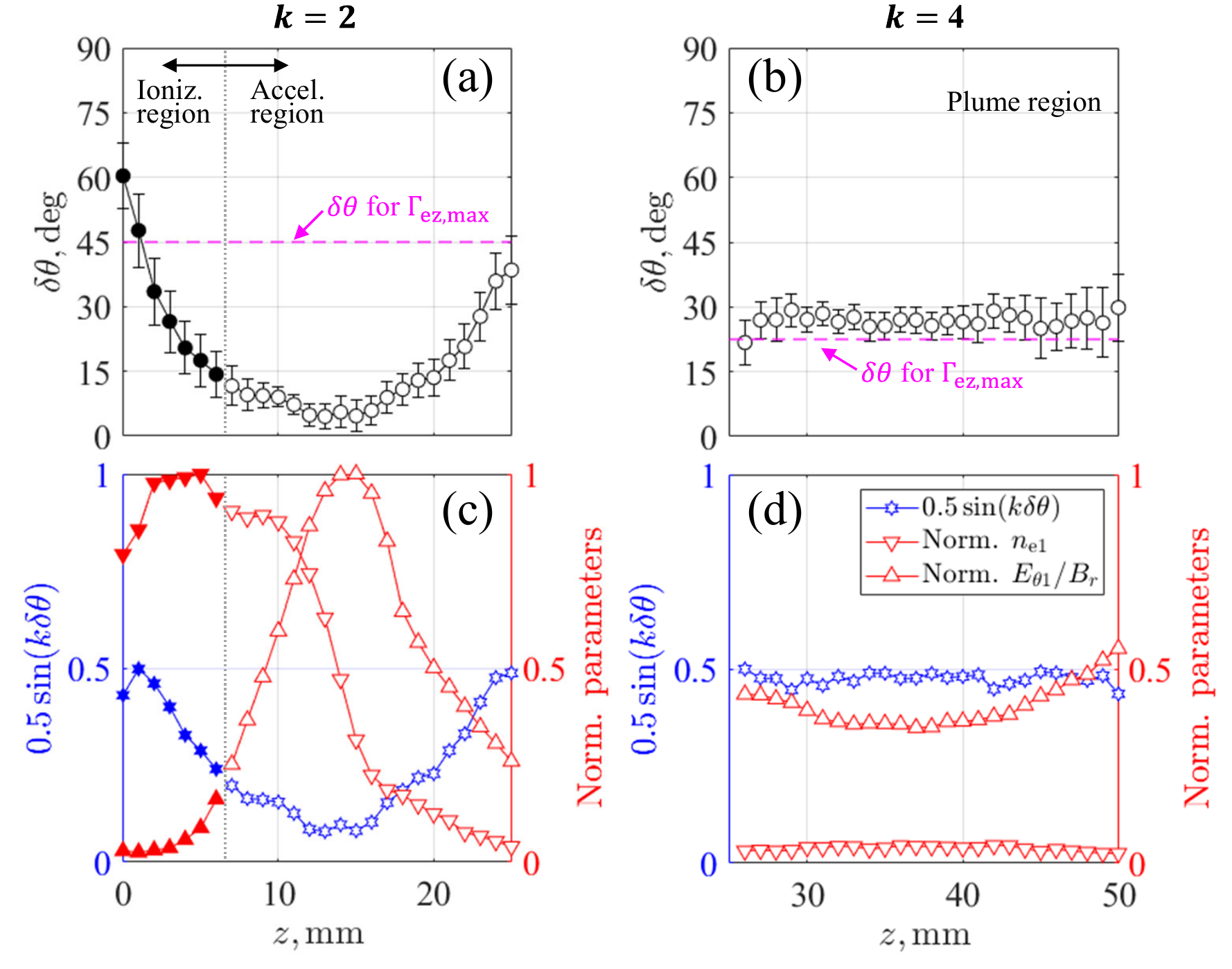}
	\caption{(a) The azimuthal phase difference $\delta \theta$ between the plasma potential and the electron density as a function of axial location are shown for $k=2$ region. The magenta dashed line shows $\delta \theta$ which maximizes $\Gamma_{\mathrm{e}z}$. (b) $\delta \theta$  for $k=4$ region. (c) The effective weight  coefficient $0.5 \sin{(k \delta \theta)}$ (blue hexagon), the normalized $n_\mathrm{e1}$ (red downward triangle), and the normalized $E_{\theta 1}/B_r$ (red upward triangle) are shown for $k=2$ region, and (d) for $k=4$ region. The normalized values are normalized by the maximum value along $z$. Note that $\Gamma_{\mathrm{e}z, E_\theta}^- = 0.5\sin{(k\delta\theta)}\cdot {n_\mathrm{e1}}\cdot\frac{E_{\theta\mathrm{1}}}{B_r}$.}
	\label{fig:delthek}
\end{figure*}

In Fig.~\ref{fig:delthek}, the phase difference $\delta \theta$ between $V_\mathrm{p}$ and $n_\mathrm{e}$ are shown for (a) $k=2$ and (b) $k=4$ regions. The parameters given in Sec.~(\ref{sec:paraset}) as a function of $z$ are shown in (c) and (d). In (a) and (c), points in the ionization region are represented with the solid markers. $E_{\theta\mathrm{1}}$ and $B_r$ are combined as $E_{\theta\mathrm{1}}/B_r$ in order to reduce one parameter and to represent the maximum $E_\theta \times B_r$ drift speed. $n_{\mathrm{e1}}$ and $E_{\theta\mathrm{1}}/B_r$ are normalized by the maximum value in the axial direction. 

First, note that the phase difference $\delta \theta$ takes a positive value at every axial position, which means plasma is formed to have an electron flow toward the anode. Secondly, $\delta \theta$ stays at a nearly fixed value in the $k=4$ region, which is very close to the $\delta \theta \approx \delta \theta ^*$ which maximizes the total axial flux. It seems the self-organized plasma structure forms the path of high conductivity for axial electron transport. While entering the $k=2$ region where $B_r$ starts stronger, $\delta \theta $ varies along with $n_\mathrm{e1}$ and $E_{\theta\mathrm{1}}$, and has a mostly clear negative correlation with the drift velocity $E_{\theta\mathrm{1}}/B_r$, suggesting that $\delta \theta$ trades off the changes in the drift velocity $E_{\theta\mathrm{1}}/B_r$ in a way of conserving the axial flux. More detailed discussion is made in the next section.

\subsection{Force balance and its relation to the spatial evolution of the affecting parameters}\label{sec:forbal}
When we think of mass conservation, the flux is conserved in the axial direction because boundaries in the radial and azimuthal directions are closed. It should be noted that strict evaluation of flux conservation along $z$ direction cannot be made from the $z$-$\theta$ data because the actual electron flux is conserved through the full 3D space in the real situation. Due to the unknown radial distribution, which would not be consistent along the axial direction, no conclusive remarks on total electron current continuity can be obtained. Strict flux conservation will only be able to be confirmed by a full 3D simultaneous measurement of related plasma parameters, or 2D numerical simulation where the radial dimension is treated to be uniform. Such an approach could provide further knowledge to identify dominant frequency components on cross-field transport by various causes of $E_\theta$. Instead, in this work, we will utilize the force balance, which can show the contribution of $\Gamma_{\mathrm{e}z, E_\theta}^-$, and discuss the spatial evolution of parameters qualitatively, which still provides several useful insights.

By focusing on the components perpendicular to the magnetic field, the momentum conservation equation for electrons in the azimuthal direction with the drift-diffusion approximation gives
\begin{eqnarray}
	m_\mathrm{e} n_\mathrm{e} \nu_\mathrm{en}  {v_{\mathrm{e}\theta}} &=& - e n_\mathrm{e} E_\theta - e n_\mathrm{e} {v_{\mathrm{e}z}} B_r - e \nabla_\theta p_\mathrm{e}.
\end{eqnarray}
The first term from the left represents the resistive force $f_{\mathrm{R},\theta}$, followed by the electric force $f_{\mathrm{E},\theta}$, the magnetic force $f_{\mathrm{M},\theta}$, and the pressure force $f_{\mathrm{P},\theta}$, respectively.

\begin{figure}
	\centering
	\includegraphics[width=5.6cm]{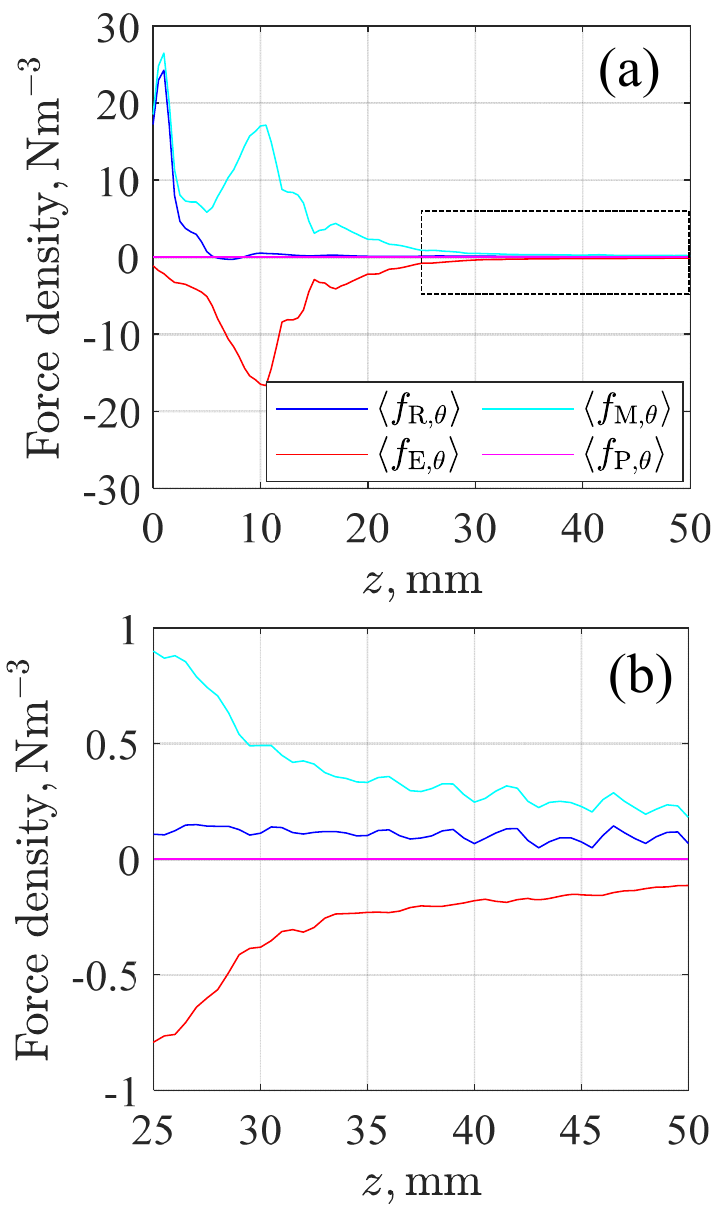}
	\caption{Force density balance in $\theta$ direction as a function of axial position (a) and the region of $z$ = 25 - 50 mm (in the dotted box) are magnified in (b).}
	\label{fig:forcbal}
\end{figure}

To calculate each force term on $z$-$\theta$ locations, first, the momentum equations in axial and azimuthal direction are solved for both velocity components (two equations and two unknowns), then, the velocities are obtained with the available quantities ($V_\mathrm{p}$, $T_\mathrm{e}$, $n_\mathrm{e}$, $B_r$ and $n_\mathrm{n}$). Finally, each force term is calculated, and their azimuthal averages as a function of the axial location are shown in Fig.~\ref{fig:forcbal}(a), and the plume region is magnified in Fig.~\ref{fig:forcbal}(b). When the forces are integrated in $\theta$, $\left \langle  f_{\mathrm{P},\theta} \right \rangle$ goes to zero, thus the pressure force is locally effective, but not globally on the net momentum balance.

From Fig.~\ref{fig:forcbal}(a), it can be seen that the resistive force $f_{\mathrm{R},\theta}$ is only considerable at near-anode, where $n_\mathrm{n}$ is high (see Fig.~\ref{fig:vpnedis}(b) top). In this region, the electric force becomes negligible ($f_{\mathrm{E},\theta}\approx 0$) and so the momentum balance is established by the resistive force $f_{\mathrm{R},\theta}$ and the magnetic force $f_{\mathrm{M},\theta}$ as seen in Fig.~\ref{fig:forcbal}(a), thus,
\begin{equation}
	\left\langle m_\mathrm{e} n_\mathrm{e} \nu_\mathrm{en}  {v_{\mathrm{e}\theta}} \right\rangle = - 0 - \left\langle e n_\mathrm{e} {v_{\mathrm{e}z}} B_r\right\rangle - 0.
\end{equation}
It basically represents $n_\mathrm{e} {v_{\mathrm{e}\theta}} = \Omega_\mathrm{H} n_\mathrm{e} {v_{\mathrm{e}z}}$, corresponding to the classical diffusion. Here,  $\Omega_\mathrm{H}(\equiv \omega_\mathrm{c}/\nu_\mathrm{en})$ is the Hall parameter where $\omega_\mathrm{c}(\equiv e B_r/m_\mathrm{e})$ is the electron cyclotron frequency. Rearranging the equation gives
\begin{equation}
	\left\langle n_\mathrm{e} {v_{\mathrm{e}z}^-} \right\rangle  \approx - \frac{m_\mathrm{e}}{e B_r^2} \left\langle  n_\mathrm{e} \nu_\mathrm{en}  E_z \right\rangle \equiv \Gamma_{\mathrm{e}z, \mathrm{cla}}^-, \label{eq:cla}
\end{equation}
thus, the axial flux is conserved by the classical collisional transport with a $1/B^2$ dependency,\cite{Chen1984} which is indeed known to be met in this region.\cite{Adam2008, Hofer2008}

On the other hand, in the acceleration and plume region where $f_{\mathrm{R},\theta}$ is negligible, the force balance becomes mainly the balance of the electric force $f_{\mathrm{E},\theta}$ and the magnetic force $f_{\mathrm{M},\theta}$ as shown in Fig.~\ref{fig:forcbal}(a) and (b), 
\begin{equation}
	0 \approx - \left\langle e n_\mathrm{e} E_\theta \right\rangle  - \left\langle e n_\mathrm{e} {v_{\mathrm{e}z}} B_r \right\rangle  - 0. \label{eq:fbal}
\end{equation}
It can be rearranged as
\begin{equation}
	\left\langle  n_\mathrm{e} {v_{\mathrm{e}z}^-} \right\rangle  \approx  \frac{\left\langle n_\mathrm{e} E_\theta \right\rangle}{B_r} \equiv \Gamma_{\mathrm{e}z, E_\theta}^- \label{eq:flux_cons},
\end{equation}
which tells that the axial net flux at these regions is conserved by the transport intrinsic to the induced $E_\theta$, $\Gamma_{\mathrm{e}z, E_\theta}^-$; thus, the classical theory is not satisfactory here, and the transport is under $1/B$ dependency (Detailed discussion on the axial mobility is given in the next section).

Consequently, the force balance shows that the phase difference between the plasma potential and density in azimuth dimension is established to balance between the magnetic force and the electric force and to conserve the axial flux. It is noteworthy that this approach is similar to that used by Morozov \etal~to determine the thermal potential along the magnetic field, where pressure and electric forces are balanced. The thermal potential $V_\mathrm{p} - T_\mathrm{e} \ln{n_\mathrm{e}} = \mathrm{constant}$, gives the relation between the potential and the density along the lines of force.\cite{Morozov1972} 

Revisiting Fig.~\ref{fig:delthek}, considering the $-z$ direction, in which electrons flow from the cathode side, the spatial evolution of the parameters can be qualitatively explained as follows. In the plume region, $\Gamma_{\mathrm{e}z}\approx$ constant due to sparse ionization. Here, because ion acceleration is fully established, $n_{\mathrm{e1}}$ and $E_{\theta\mathrm{1}}$ rarely change, so $\delta\theta$ is also maintained respectively. In the acceleration region, $\Gamma_{\mathrm{e}z}$ is still approximately constant due to sparse ionization, however, as $n_{\mathrm{e1}}$ and $E_{\theta\mathrm{1}}$ increase, $\delta\theta$ decreases. Lastly, in the ionization region, $\Gamma_{\mathrm{e}z}$ increases due to the ion generation. Here, as $n_{\mathrm{e1}}$ saturates and  $E_{\theta\mathrm{1}}$ decreases, $\delta\theta$ increases accordingly.

\subsection{Regional evolution of the effective axial mobility of $\Gamma_{\mathrm{e}z, E_\theta}^-$} \label{sec:mobil}
Lastly, we look into the effective axial mobility of $\Gamma_{\mathrm{e}z, E_\theta}^-$. For this purpose, one needs to re-organize the equation as a function of the axial electric field. By introducing the coefficient for the phase relation $\alpha \equiv 0.5{\sin( k\delta \theta)}$, the electron density inhomogeneity level $\beta \equiv n_\mathrm{e1}/n_\mathrm{e0}$ and the E-field ratio $\gamma \equiv E_{\theta1}/{\left \langle E_z \right \rangle}$, Eq.~(\ref{eq:gam_ez_}) can be further adjusted to obtain the effective axial mobility coefficient as follows:
\begin{eqnarray}
	\Gamma_{\mathrm{e}z, E_\theta}^- &=& \frac{ \alpha \beta \gamma}{ B_r}  n_\mathrm{e0}{\left \langle E_z \right \rangle}  = \frac{\kappa}{B_r}n_\mathrm{e0}{\left \langle E_z \right \rangle}, \label{eq:flux_cor}
\end{eqnarray}
where $\kappa\equiv \alpha \beta \gamma$. Note that the transport induced by $E_\theta$ yields the mobility equivalent scaled by $1/B$, and its coefficient of proportionality $\kappa$ is determined by the three parameters; $\alpha$ , $\beta$, and $\gamma$. Thus, all three parameters become important in determining the transport. $\alpha$, which is the effective weight coefficient from the azimuthal phase difference $\delta \theta$ between $n_\mathrm{e}$ and $V_\mathrm{p}$, is enhanced as $\delta \theta \rightarrow \pi/2k$. This parameter shows the importance of taking into account the full azimuthal distributions of $n_\mathrm{e}$ and $V_\mathrm{p}$. $\beta$ and $\gamma$, which are proportional to the inhomogeneity of $n_\mathrm{e}$ and $V_\mathrm{p}$ respectively, stress that the cross-field transport can increase quadratically due to the inhomogeneity. This indicates the importance of suppressing inhomogeneity. For transport by the equilibrium distribution, this could be achieved by improving uniformity of operation parameters (i.e.~propellant supply method, magnetic circuit design, and anode shape), and for transport from instabilities, by suppressing the growth. This multi parameter dependency somewhat explains why various plasma conditions, such as those existing in different plasma devices or regimes or the generation of $E_\theta$ by different causes, have led to various different anomalous mobility coefficients\cite{Janes1966, Mouthaan1968, Hirakawa1995} as well as variation of the coefficient observed long the channel axis in Hall thrusters\cite{Hagelaar2003,Adam2008,Hofer2008,Bak2019}.

The expression of the coefficient $\alpha$ infers that even if the plasma is formed in a way maximizing the correlation effect, $\sin( k\delta \theta) \rightarrow 1$, $\alpha$ cannot be greater than 0.5, thus $\alpha \leq 0.5$. Also, as the plasma density cannot be negative, $\beta < 1$ is satisfied. Lastly, considering the scale length in $\theta$ and $z$ direction, $R_\theta > R_z$, for a variation of the plasma potential $\delta V_\mathrm{p}$, $\delta V_\mathrm{p}/R_\theta < \delta V_\mathrm{p}/R_z$ holds, so, $\gamma  < 1$ is met. Thus, the final condition $\kappa < 0.5$ is obtained. This limitation is in accordance with the fact that the maximum values of $\kappa$ reported in multiple experiments were smaller than 0.5.\cite{Spitzer1960,Janes1966}

\begin{figure}
	\centering
	\includegraphics[width=6.5 cm]{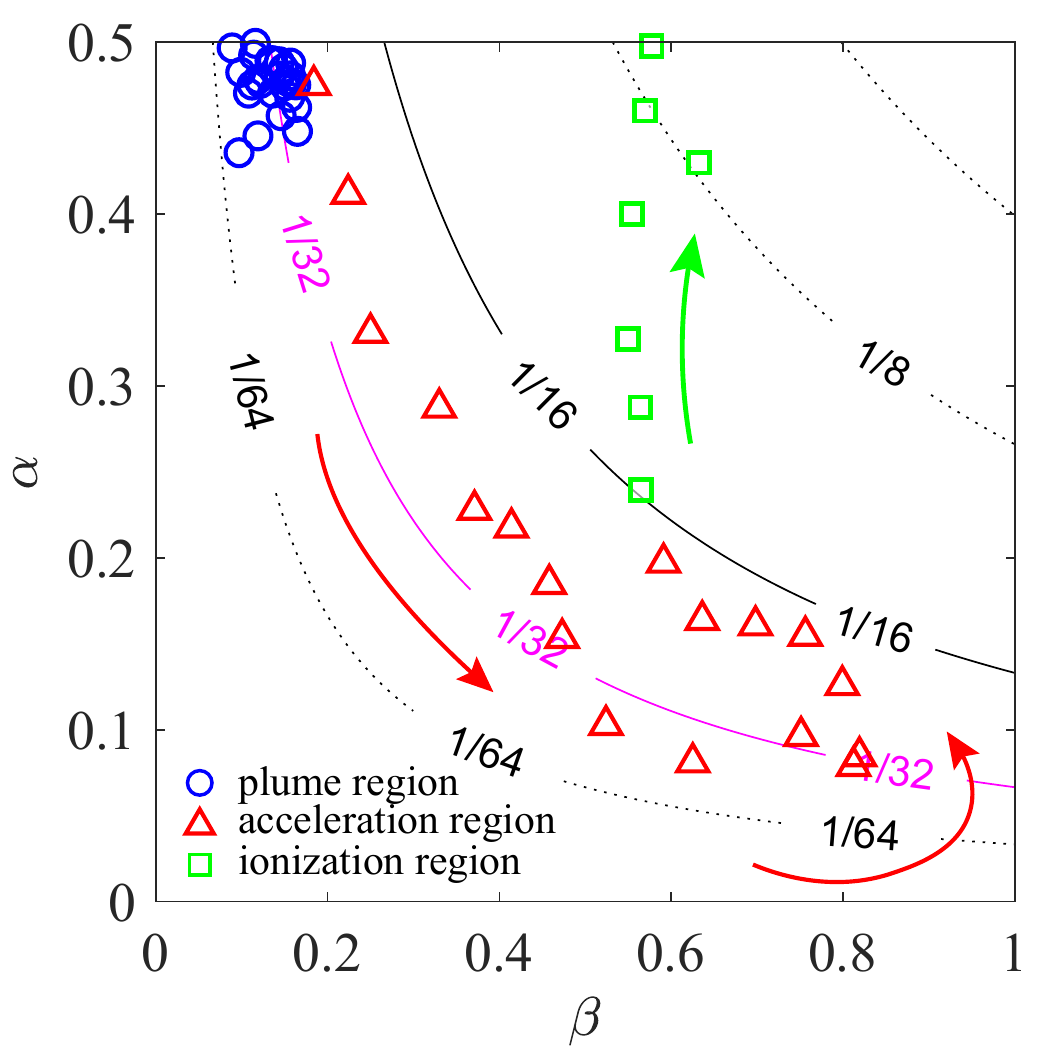}
	\caption{The contour map of the coefficient $\kappa$ as a function of the phase relation $\alpha(\equiv 0.5{\sin( k\delta \theta)})$, the plasma density inhomogeneity level $\beta(\equiv n_\mathrm{e1}/n_\mathrm{e0})$. The blue circles are those of the plume region ($k=4$), the red triangles are those of the acceleration region ($k=2$), and the green squares are those of the ionization region ($k=2$) where the classical diffusion starts to be non-negligible closer to the near-anode region. The arrows show the electron flow direction from the cathode side to the anode.}
	\label{fig:alphaplot}
\end{figure}

The contour map of the mobility coefficient $\kappa$ as a function of $\alpha$ and $\beta$ is shown in Fig. \ref{fig:alphaplot}. Please note that because $\gamma$ in this work was relatively constant throughout the resolved region at $0.47 \pm 0.17$, it is not represented as an axis but is included in the $\kappa$ contours by multiplying $0.47$. 

It is interesting to see that the blue circles from the plume region and the red triangles from the acceleration region are closely distributed along the specific contour line of $1/32$. These regions were where the electric force $f_{\mathrm{E},\theta}$ and the magnetic force $f_{\mathrm{M},\theta}$ are balanced so electrons were carried by $\Gamma_{\mathrm{e}z, E_\theta}^-$. This suggests plasma seems self-organized to maintain the mobility $\kappa$ of $\Gamma_{\mathrm{e}z, E_\theta}^-$ throughout the plume and acceleration regions until other force terms become involved; in this case, the resistive force as entering the ionization region at $z<7$ mm where the main mechanism of cross-field transport becomes the classical diffusion $\Gamma_{\mathrm{e}z, \mathrm{cla}}^-$.

The value $1/32$ is within the commonly known uncertainty of a factor three of the Bohm coefficient $\kappa_\mathrm{B}=1/16$.\cite{Bohm1949} However, please note that this work is not to argue the neutral inhomogeneity is the likely reason for the Bohm transport of turbulent transport. This is because, due to the nature of time-averaged data, the present work does not take into account any cross-field transport induced by the time-varying physics. Rather, we note that the $1/B$ scaling is not limited to the Bohm or time fluctuating transport. It would be more reasonable to say that the obtained $\kappa$ is one specific case of the cross-field transport by the induced equilibrium $E_\theta$ from the given input neutral inhomogeneity. However, the spatial evolution of affecting parameters and its qualitative explanation over the distinct regions in this work are expected to expand our understanding that had been limited to a limited spatial location. A similar analysis approach could be applied to situations where other causes induce $ E_\theta$ in different spatiotemporal scales, such as spokes\cite{Ellison2012, Hecimovic2018, Kawashima2018} or electron drift instability\cite{Lafleur2017, Boeuf2018}, and it would be interesting to see how the $\kappa$ coefficients in other spatiotemporal scales evolve along with other plasma parameters.

\section{Summary}\label{sec:sum}
In summary, we investigated the spatial evolution of the parameters that determine the axial electron flux by the induced $E_\theta$ under inhomogenous neutral supply. 

A fast Fourier transform analysis of the plasma structure reveals that the wavenumber $k$ of the azimuthal plasma structure increases from $k=2$, which is the input condition, to $k=4$ in the plume region. It is also found that the total axial flux caused by the azimuthal electric field was mainly resulting from the structures of the dominant Fourier components. Regarding the increased wavenumber of the plasma structure at the downstream region, it is hypothesized that the plasma structure may have evolved to a finer structure from bulk ion concentrations.

The phase difference between the plasma potential and density in azimuth reveals that, in the plume region, plasma structure is formed in a way to maximize the axial flux, and starts varying as other plasma parameters change toward the anode. The spatial evolution of parameters involved in the axial electron transport by the $E_\theta$ is qualitatively explained from the force balance.

The effective axial mobility in the form of $\kappa/B_r$ shows that the proportional coefficient $\kappa$ is closely maintained near the 1/32 line in the regions where the magnetic force and the electric force balance, and starts deviating in the near anode region where the resistive force starts to be non-negligible and so the classical theory becomes appreciable. 

\begin{acknowledgments}
	This work was supported by JSPS KAKENHI Grant Number JP18J14592 and JP20H02346. The authors would like to acknowledge the Princeton-University of Tokyo Strategic Partnership and Masaaki Yamada for enabling the collaboration.
\end{acknowledgments}

\section*{Data availability}
The data that support the findings of this study are available from the corresponding author upon reasonable request.

\appendix
\section{Fitting of the azimuthal distribution}\label{sec:app}
\begin{figure*}
	\centering
	\includegraphics[width=0.9\linewidth]{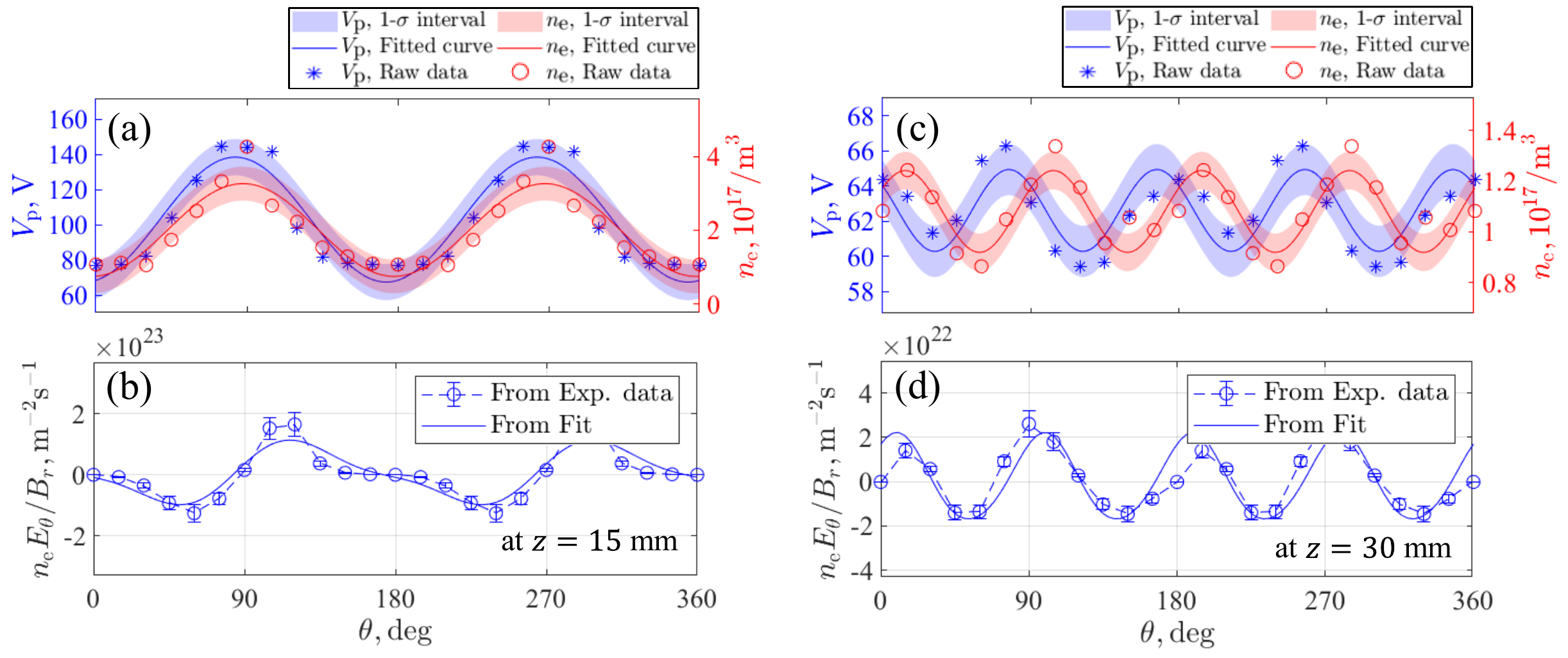}
	\caption{Fitting result examples at $z=15$ and $z=30$ mm. (a,c) The azimuthal distribution of plasma potential $V_\mathrm{p}$ and electron density $n_\mathrm{e}$, where points are experimental data and solid lines are the fitted curves with $1$-$\sigma$ confidence band of the fit shown as the shaded region, (b,d) Comparison of $n_\mathrm{e} E_\theta / B_r$ distribution between experimental data and fitting.}
	\label{fig:fitting}
\end{figure*}

Fitting of azimuthal distribution of plasma potential and density is done in the form of $f(\theta) = a + b \sin(k(\theta+c))$ from $z=0$ mm to $z=50$ mm every 1 mm. Examples of fitting analysis at $z=15$ mm and $z=30$ mm are shown in Fig.~\ref{fig:fitting}. These locations are chosen as the representative examples of the regions of the wave number $k =2$ and $k=4$. 

In Fig.~\ref{fig:fitting}(a), the azimuthal distribution of plasma potential $V_\mathrm{p}$ and electron density $n_\mathrm{e}$ (experimental data) for $z=15$ mm are shown as the blue and red points. The corresponding solid color lines show the fitted curves together with $1$-$\sigma$ confidence band as the shaded region. Fig.~\ref{fig:fitting}(c) is those of $z=30$ mm. Fig.~\ref{fig:fitting}(b) shows a comparison of $n_\mathrm{e} E_\theta / B_r$ distribution between experimental data and fitting, and Fig.~\ref{fig:fitting}(d) shows those of $z=30$ mm. 

The result of every 5 mm are given in Table~\ref{table:fits}. In the table, 1-$\sigma$ confidence bounds of the fitting are shown as errors for each fitting coefficient. Relative root mean square error (RRMSE) is defined as 
\begin{equation}
	\text{RRMSE} = \dfrac{\sqrt{\dfrac{1}{N} \sum\limits_{i=1}^{N} \left( X_i - f_i \right)^2  }}{\dfrac{1}{N} \sum\limits_{i=1}^{N} f_i} \times 100\%,
\end{equation}
where $X_i$ is the $i$th experimental data and $f_i$ is the $i$th value from the fitted function.

\begin{table*}
	\centering
	\caption{Fitting results of azimuthal distribution of the plasma potential $V_\mathrm{p}$ and the electron density $n_\mathrm{e}$ at multiple axial locations. The fit function is given as $f(\theta) = a + b \sin(k(\theta+c))$, where $k$ is given from the fast Fourier transform.}
	\label{table:fits}
	\ra{1.2}
	\setlength{\dashlinedash}{0.2pt}
	\setlength{\dashlinegap}{4.5pt}
	\setlength{\arrayrulewidth}{0.2pt}
	\begin{threeparttable}
		\begin{tabular}{@{}r rrrr c rrrr c r@{}}
			\toprule
			& \multicolumn{4}{c}{$V_\mathrm{p}$} & \phantom{a}& \multicolumn{4}{c}{$n_\mathrm{e}$} & \phantom{a}& \\
			\cmidrule{2-5} \cmidrule{7-10}
			$z$, mm& \multicolumn{1}{c}{$a$} & \multicolumn{1}{c}{$b$}  & \multicolumn{1}{c}{$c$}  & \thead[r]{RRMSE\\ \%}&  &\multicolumn{1}{c}{$a$} & \multicolumn{1}{c}{$b$}  & \multicolumn{1}{c}{$c$}   &  \thead[r]{RRMSE\\ \%} & & \multicolumn{1}{c}{$\delta \theta$, deg\tnote{a}}\\
			\midrule\midrule[0.1em]
			$(k=2)$ \\
			0  & 154.2$\pm$0.2   &   1.2$\pm$0.3     & 173.3$\pm$7.5    &      1     & &   5.00$\pm$0.15   &   3.16$\pm$0.21    & 113.0$\pm$ 1.9 &    15  & &   60.3$\pm$7.7 \\ 
			5  & 150.9$\pm$0.8   &   5.8$\pm$1.1     & 145.2$\pm$5.8    &      3     & &   7.06$\pm$0.14   &   3.99$\pm$0.20    & 127.6$\pm$ 1.5 &    10  & &   17.5$\pm$6.0 \\ 
			10  & 130.8$\pm$1.4   &  30.6$\pm$1.9     & 144.3$\pm$1.9    &      5     & &   4.62$\pm$0.11   &   3.49$\pm$0.15    & 135.3$\pm$ 1.3 &    11  & &    9.0$\pm$2.3 \\ 
			15  & 103.0$\pm$2.0   &  35.5$\pm$2.7     & 141.9$\pm$2.3    &      9     & &   2.01$\pm$0.09   &   1.26$\pm$0.12    & 137.3$\pm$ 2.8 &    21  & &    4.7$\pm$3.6 \\ 
			20  &  80.6$\pm$1.0   &  12.1$\pm$1.3     & 149.2$\pm$3.2    &      6     & &   1.34$\pm$0.03   &   0.50$\pm$0.05    & 135.6$\pm$ 2.9 &    13  & &   13.6$\pm$4.3 \\ 
			$(k=4)$ \\
			25  &  70.0$\pm$0.7   &   3.8$\pm$1.0     &  35.3$\pm$3.9    &      5     & &   1.11$\pm$0.03   &   0.10$\pm$0.04    &  11.2$\pm$ 5.9 &    13  & &   24.2$\pm$7.0 \\ 
			30  &  62.6$\pm$0.3   &   2.3$\pm$0.4     &  35.8$\pm$2.4    &      2     & &   1.08$\pm$0.01   &   0.16$\pm$0.02    &   8.7$\pm$ 1.7 &     6  & &   27.1$\pm$3.0 \\ 
			35  &  58.3$\pm$0.1   &   1.5$\pm$0.2     &  34.2$\pm$1.8    &      1     & &   1.09$\pm$0.02   &   0.16$\pm$0.03    &   8.6$\pm$ 2.6 &     9  & &   25.6$\pm$3.2 \\ 
			40  &  55.4$\pm$0.1   &   1.0$\pm$0.1     &  33.2$\pm$1.9    &      1     & &   1.09$\pm$0.03   &   0.16$\pm$0.04    &   6.6$\pm$ 3.3 &    12  & &   26.5$\pm$3.9 \\ 
			45  &  53.6$\pm$0.1   &   0.8$\pm$0.1     &  33.1$\pm$2.1    &      1     & &   1.07$\pm$0.04   &   0.12$\pm$0.06    &   8.1$\pm$ 6.6 &    18  & &   25.0$\pm$7.0 \\ 
			50  &  52.1$\pm$0.1   &   0.7$\pm$0.1     &  34.8$\pm$2.2    &      1     & &   1.04$\pm$0.04   &   0.10$\pm$0.05    &   5.0$\pm$ 7.5 &    18  & &   29.8$\pm$7.8 \\ 
			\bottomrule
		\end{tabular}
		\begin{tablenotes}
			\item[a]$\Gamma_{\mathrm{e}z, E_\theta}^-$ is maximized at $\delta\theta^* \equiv \frac{\pi}{2k}$. For $k=2$,  $\delta\theta^* = 45$ deg. For $k=4$, $\delta\theta^* = 22.5$ deg.        
		\end{tablenotes}
	\end{threeparttable}
\end{table*}

\section*{References}
\bibliography{reference}

\end{document}